\documentclass[cpp,fleqn]{w-art}
\usepackage{times}
\usepackage{w-thm}
\theoremstyle{plain}

\theoremstyle{definition}

\usepackage[]{graphicx}
\chardef\bslash=`\\ 

\begin{document}
\DOIsuffix{theDOIsuffix}
\Volume{42}
\Issue{1}
\Month{01}
\Year{2003}
\pagespan{3}{}
\Receiveddate{27 October 2006}
\Reviseddate{}
\Accepteddate{}
\Dateposted{}
\subjclass[pacs]{}



\title[Crystallization in mass-asymmetric electron-hole bilayers]{Crystallization in mass-asymmetric electron-hole bilayers}
                   

\address[\inst{1}]{Christian-Albrechts-Universit\"at zu Kiel, Institut
f\"ur Theoretische Physik und Astrophysik, Leibnizstrasse 15,
24098 Kiel, Germany}
\address[\inst{2}]{Universit\"at Rostock, Institut f\"ur Physik,
Universit\"atsplatz 3, 18051 Rostock, Germany}
\address[\inst{3}]{Institute of Spectroscopy RAS, Moscow region,
Troitsk, 142190, Russia}

\author[P. Ludwig]{P.~Ludwig\footnote{Corresponding author: e-mail: {\sf ludwig@theo-physik.uni-kiel.de}, Phone: +49\,(0)431\,880\,4732, Fax: +49\,(0)431\,880\,4094}\inst{1,2}} 
\author[A. Filinov]{A.~Filinov\inst{1,3}}
\author[Yu.E. Lozovik]{Yu.E~Lozovik\inst{3}}
\author[H. Stolz]{H.~Stolz\inst{2}}
\author[M. Bonitz]{M.~Bonitz\inst{1}}

\begin{abstract}
We consider a \textit{mass-asymmetric} electron and hole bilayer. Electron and hole
Coulomb correlations and electron and hole quantum effects are treated on first princles 
by path integral Monte Carlo methods.
For a fixed layer separation we vary the mass ratio $M$ of holes and electrons between $1$ 
and $100$ and analyze the structural changes in the system. While, for the chosen density,
the electrons are in a nearly homogeneous state, the hole arrangement changes from 
homogeneous to localized, with increasing $M$ which is verified for both, mesoscopic 
bilayers in a parabolic trap and for a macroscopic system.
 
\end{abstract}
\maketitle                   

\section{Introduction}

Strongly correlated Coulomb systems are of growing interest in many fields, including 
plasmas and condensed matter, see e.g. \cite{kalman98} for an overview. In particular,
Wigner crystal formation is one of the most prominent correlation phenomena observed in 
ultracold ions \cite{ions}, dusty plasmas \cite{dust,bonitz-etal.06prl}, quantum dots, e.g. \cite{AF01,Nature06} and other confined (non-neutral) systems.
Recently crystal formation in two-component (neutral) quantum plasmas was demonstrated by simulations \cite{MB06} confirming early predictions of hole crystallization in semiconductors 
by Halperin and Rice \cite{halperin}, Abrikosov \cite{abrikosov} and others. Interstingly, 
this is essentially the same physical phenomenon as crystallization of nuclei in White Dwarf 
stars \cite{segretain}.

A different type of two-component system, standing in between the neutral and non-neutral 
Coulomb systems, are bilayer containing spatially separated positive and negative 
charges which are most easily to realize in semiconductors by means of doping (electron-hole 
bilayers). These systems are of high interest because the strength of the correlations can be 
tuned by varying the layer separation $d$. The interplay of intra-layer and inter-layer 
correlations in classical bilayers has been studied in detail for macroscopic, e.g. 
\cite{Donko} and mesoscopic \cite{PL03,PLdip} systems. Quantum bilayers have been treated much 
less, see e.g.  \cite{Rapisarda,AF03,Peeters02_PRB65,Peeters02_PRB66} and are much poorer 
understood. In particular, most investigations have considered symmetric bilayers, where the 
hole to electron mass ratio $M=m_h/m_e$ equals one.
However, the typical mass ratio in semiconductors is on the order of $M=3\dots 10$, and even 
exotic materials exist where $M$ reaches $40$ \cite{Wachter} or even higher values.

For this reason, in this paper we concentrate on the effect of the mass ratio on crystal 
formation in quantum electron-hole bilayers. Varying $M$ from $1$ to $100$ at low temperature 
and high density, we can tune the hole behavior from delocalized (quantum) to localized (quasi-classical) while the electrons remain delocalized all the time. 
As was recently observed for bulk semiconductors \cite{MB06}, holes undergo a phase transition to a crystalline state if the mass ratio exceeds a critical value of $M_{cr}\approx 80$. Here, we 
extend this analysis to bilayers where $M_{cr}$ depends on $d$ and the in-layer 
particle density. To reduce the complexity of the problem, here we will keep $d$ fixed. 
The complicated overlap of correlation and quantum effects of both, electrons and holes, is 
fully taken care of by performing first-principle path integral simulations. We present results 
for two types of e-h bilayers: a mesocsopic system of $N=36$ particles in a parabolic trap and 
for a macroscopic system of the same density.

\section{Model and Parameters}\label{model}
The physical realization of the mass asymmetric bilayers considered here
can be a system of two coupled quantum wells filled with electrons and holes, respectively. An additional in-plane potential can produce the lateral confinement of the carriers leading to a system of two coupled quantum dots. Recently, we have analyzed in detail a possible realization of a parabolic in-plane potential using the idea of the quantum Stark confinement~\cite{PL06}. An inhomogenous electric field applied perpendicular to the QW plane changes the energy of a particle in the quantum well because the penetration of a particle inside the barrier material depends on the strength of the electric field. For example, in GaAs and ZnSe based QW one can achieve harmonic trap frequencies from $1$~GHz to $1$~THz for typical electric field strengths of $10-20$~kV/cm.

In this paper, we approximate two coupled QWs by a model of two vertically separated 2D layers populated with $N_e$ electrons and $N_h$ holes (we consider the case $N_e=N_h=N/2$). The charges interact via the Coulomb potential. The underlying Hamiltonian is well defined and is of practical importance for semiconductor heterostructures
\begin{eqnarray}
\hat H = \hat H_{e} + \hat H_{h} + \sum\limits_{i=1}^{N}\sum\limits_{j=i+1}^{N} \frac{e_i e_j }{\varepsilon \sqrt{({\bf r}_i- {\bf r}_j)^2+(z_i -z_j)^2}} , \quad
\hat H_a = \sum\limits_{i=1}^{N_{a}} \left( -\frac{\hbar^2}{2 m^*_{a}} \nabla_{{\bf r}_i}^2 + \frac{m^*_a}{2}\omega_a^2 r_i^2 \right),
\label{Ham}
\end{eqnarray}
where the electrons (e) are confined to the plane $z=0$ and the holes (h) to the plane $z=d$; also ${\bf r}_i$ and ${\bf r}_j$ are the in-plane 2D radius vectors 
describing the particle coordinates in each layer.
In the following all lenghts will be given in units of the effective Bohr radius $a_B=\hbar^2\epsilon/m^*_e e^2$. For example, for GaAs and ZnSe quantum wells
this results in the length units $a_B(GaAs)=9.98$~nm and $a_B(ZnSe)=3.07$~nm. Energies and temperatures are measured in Hartree units:
$1Ha(GaAs)=11.47 meV (133.1~K) $ and $1Ha(ZnSe)=53.93 meV (625.8~K)$.

For the mesoscopic trapped system the density is controlled by the harmonic trap frequency (we use $m^*_e\omega_e^2=m^*_h\omega_h^2$) and is characterized by the coupling parameter $\lambda=(e^2/\epsilon l_0)/(\hbar \omega_e)=l_0/a_B$ with $l_0^2=\hbar/m^*_e \omega_e$. In this case, the coupling parameter for the holes is related to the electron coupling as $\lambda_h=\lambda (m^*_h/m^*_e)^{3/4}$.
Also for Coulomb systems in a parabolic trap one can find the following usefull relations. 
For two classical particles in a parabolic trap their separation distance $r_0$ in the ground state is given by: $e^2/\epsilon r_0=m_e\omega_e^2 r_0^2/2$. Now we can define the density parameter $\tilde r_s$ (in analogy to the Brueckner parameter $r_s=\langle r \rangle /a_B$ for macroscopic systems) as follows: $\tilde r_s = r_0/a_B=(2e^2/\epsilon m_e\omega_e^2)^{1/3}/a_B=2^{1/3} \lambda^{4/3}$. We will use this formula to obtain approximate relations between the densities in the mesoscopic and macroscopic system by relating
$\lambda \leftrightarrow \tilde r_s \leftrightarrow r_s$.

\subsection{Numerical details}

To solve the problem of $N$ interacting particles described by the Hamiltonian~(\ref{Ham}) we use the path integral Monte Carlo (PIMC) method.
The applied PIMC simulation technique was described in detail in 
Ref.~\cite{numbook06}. The effective interaction potentials used in the
expressions for the high-temperature pair density matrices were obtained by using the matrix squaring technique~\cite{storer, Cep95}.

One of the main obstacles that limit applicability of the PIMC method
for systems of particles obeing Fermi statistics is the so called {\em Fermion sign problem}. Without additional approximations the direct fermionic PIMC simulations are only limited to  problems where the degeneracy is not very high. This, certainly depends on the physical situation and is related to the particle density, interaction strength and temperature. Full inclusion of the quantum exchange
effects for the number of particles considered here, i.e $N_{e(h)}\approx 36 - 64$,
will not be possible without neglecting the spin statistics and permutations
in the electron and hole subsystems.
However, direct comparison of the PIMC simulations without spin~\cite{HilkoCords} with the results of Ref.~\cite{egger99} which include spin effects show, that the errors introduced by neglecting the spin statistics are of the order of few percents and are completely negligible for $\lambda \ge 10$. The considered here electron densities, i.e $r_s \approx 18$, are sufficiently low (for the holes the corresponding parameter $r_{s}^{(h)}$ is even larger due to their larger mass) and the dominant effect for the interparticle correlations (and in particular for the holes) are driven mainly by the strength of the Coulomb interaction and not by quantum statistics effects.
Also, there is no doubt that the spin will have a negligible effect on the localized states of the holes when they form a 2D lattice. Hence, we expect, that the solid-liquid transition investigated in this paper will not be sensitive to the particle spin. Nevertheless, the question about the true ground state of the electron liquid (i.e. spin polarized or unpolarized), just after the solid-liquid transition is currently under active discussion~\cite{Spin_order} and requires further
investigation.

In the simulations presented below we assume that the electrons and holes
can reach thermal equilibrium and are cooled down by using, e.g. $^3He/^4He$ dilution refrigeration  to a temperature of $k_BT = 1/3000$~Ha. For ZnSe (GaAs) this corresponds to an absolute value 
of $T=208.6 (44.37)$~mK. At these low temperatures thermal fluctuations are negligible and 
the system is practically in the ground state. 

In the PIMC representation of the density matrix applied in our simulations \cite{numbook06}, 
we have used $256$ (in some cases 128) beads (high temperature factors). This was sufficient to reache convergence for the full energy better than $1 \%$ and an even better accuracy for the pair distribution functions. 
The use of such a moderate number of beads was only possible by using
pre-computed tables of the pair density matrices for all types of Coulomb interactions, i.e. for the intra-layer and inter-layer interaction terms in the hamiltonian~(\ref{Ham}), and for the external parabolic confinement for both electrons and holes. 
To reduce the enormous computational effort for a simulation of fermions, here we used Boltzmann statistics for both electrons and holes, and the spin effects are omitted. For reasons discussed above we expect that this will not influence the results of this paper significantly. 
  
Both layers are treated as pure 2D layers of zero thickness. 
Considering that the thickness of real physical QWs is of the order 
of few Bohr radii, this approximation seems to be reasonable for 
the range of densities considered here, i.e. $r_s=\langle r \rangle /a_B \geq 10$,
and an inter-layer distance of $d=20 a_B$. [The case when $d$ and $\langle r \rangle$ 
become comparable to the well width would require essentially more computationally costly 
3D simulations and inclusion additional terms related to the QW potential in the 
hamiltonian~(\ref{Ham})]. 
For quite narrow QWs with a thickness of about $1 a_B$ and less, the addiabatic
approximation can be succesfully used with the 2D hamiltonian~(\ref{Ham}) with slightly changed interaction terms (see Ref.~\cite{fil_prb}).

For the chosen inter-layer distance, $d=20 a_B$, our system represents 
essentiall a 3D structure, as the intra-layer and inter-layer correlations are on the same length and energy scales. For small ratios $d/r_s \ll 1$ the system approaches the  single layer limit,
whereas for $d/r_s \gg 1$, it behaves like two uncoupled layers. Also, for $d=20 a_B$ we can completely neglect the inter-layer tunneling, and the life time of electrons and holes
can reach a few microseconds which is much larger than their equilibration time.

The mass ratio of the electrons and holes is varied in the range of 
$1 \leq m^*_h/m^*_e\leq 100$ which covers practically all semiconductor materials. In our simulations we have found that the initial equilibration time needed to bring the system from an initial randomly chosen configuration to the thermodynamic one,
depends on the electron-hole mass ratio and the strength of the external confinement. Usually we skip the first $10\,000 - 100\,000$ MC-steps and only then start to accumulate thermodynamic averages. 
 
\subsection{Calculated quantities} 
\subsubsection{Pair and radial distribution functions}
The physically relevant quantities to investigate a phase transition
are the {\em radial}, $n(r)$, and {\em pair} distribution function, $g(r)$.
Both functions are a good probe of the short and long-range order in the system and yield information on the importance of correlation effects.
In Statistical Mechanics these quantities are given by the expressions
\begin{equation}
g_{ab}(r) = \frac{1}{N_{a} N_b}\sum\limits_{i=1}^{N_a} \sum\limits_{j=1}^{N_b} \langle \delta( \vert \mathbf{r}_{ij} \vert -r)\rangle, \quad
n_{a(b)}(r)=\frac{1}{N_{a(b)}} \sum\limits_{i=1}^{N_{a(b)}} \langle \delta(\vert \mathbf{r}_{i} \vert -r_0)\rangle,
\end{equation}
where $a$ and $b$ are two particle species, $r_0$ is the reference point for the radial density (e.g the center of the parabolic potential as used here), and $\langle \ldots \rangle$ denotes the thermodynamic average. In the PIMC approach 
the averaging is performed with the $N-$particle density matrix, i.e
\begin{equation}
\langle \ldots \rangle =\frac 1 Z \int \int d\mathbf r_1 d\mathbf r_2 \ldots d\mathbf r_N
\; (\ldots)\; \rho(\mathbf r_1,\mathbf r_2,\ldots,\mathbf r_N; \beta).
\end{equation}
After the high-temperature decomposition this integral includes also additional integrations over the particle coordinates on the intermediate ``time-slices'' and, as a result,
the particle images on each time slice also contribute to the distribution function
which significantly improves the convergence of the simulations.

\subsubsection{Lindemann parameter}
One of the criteria to investigate structural
phase transitions (e.g., solid-liquid phase transition) was proposed by Lindemann~\cite{lind},
who used vibration of atoms in the crystal to explain the melting transition.
The average amplitude of thermal vibrations increases with
temperature of the solid.
At some point the amplitude of the vibrations becomes so large that the atoms start to occupy
the space of their nearest neighbors and disturb them, and the melting process is initiated.
According to Lindemann, the melting might be expected when
the root mean vibration amplitude $\sqrt{\langle \delta u^2 \rangle/a^2}$ exceeds a certain threshold value ($\langle \delta u^2 \rangle$ is the particle fluctuation from a lattice site, $a=1/\sqrt{\pi n}$, $n$ is the density).
Namely, when the amplitude reaches at least $10\%$ of the nearest neighbor distance,
this quantity exhibits a rapid growth when the temperature becomes close
to the melting temperature of the solid phase. While for 3D systems this criterion can be successfully used, in 2D this quantity shows a logarithmic divergence, $\ln(L/a)$, with the increase of the system size $L$. 
Instead, to indicate the phase transiton from a liquid to a crystal, in 2D, one should apply the modified Lindemann criterion and use the relative distance fluctuations~\cite{loz_ur}
\begin{eqnarray}\label{reldistfluct}
u^{ab}_r = \frac{1}{N_{a} N_b}\sum\limits_{i=1}^{N_a} \sum\limits_{j=1}^{N_b} \sqrt{\frac{\left\langle r_{ij}^2 \right\rangle}{\left\langle r_{ij} \right\rangle^2}-1},
\end{eqnarray}
where $r_{ij}$ is the distance between the particles $i$ and $j$.
To reduce the effect of particle diffusion through the cluster (in a finite system)
or through the simulation cell (for a macroscopic system), 
which leads to very slow convergence with the increase of the system size, 
in the calculation of (\ref{reldistfluct}) we have performed partial averaging over $1\,000$ MC-steps (one block). After the current block has been completed we proceed to a new one and the MC averaging was repeated for the next $1\,000$ MC-steps. The difference
in the fluctuations measured from block to block can  characterize the ordering in the system and is more effective for large systems.

\subsubsection{Nature of the phase transition in 2D systems}

Strictly speaking, in classical macroscopic 2D systems at $T\neq 0$ a true crystal state does not exist. The absence of off-diagonal long range order in the system
manifests itself in the existance of two disordered phases characterized by different
asymptotic behavours of the pair correlation function $g(r,r')$. The system undergoes a transition at a finite temperature $T_{KT}$ (Kosterlitz-Thouless transition) when 
the asymptotes $g(r,r')\vert_{\vert \mathbf r - \mathbf r' \vert \rightarrow \infty}$ changes from
\begin{equation}
g(r,r')\approx\frac{\exp(-\vert \mathbf r - \mathbf r' \vert/\xi(T))}{\vert \mathbf r - \mathbf r' \vert^{\alpha(T)}} \quad (T\geq T_{KT}) \quad \text{to} \quad g(r,r')\approx\frac{1}{\vert \mathbf r - \mathbf r' \vert^{\alpha'(T)}} \quad (T < T_{KT}).
\label{KT}
\end{equation}
The important question of the relevence of the standard Kosterlitz-Thouless theory also for 2D {\em quantum} systems has been disscussed in Ref.~\cite{Akopov} for the two-dimensional XY model. A generalization for Coulomb systems is subject of ongoing work~\cite{Loz06}. Concerning the interpretation of the results of the present publication
we indeed find a abrupt transition in the decay of the maxima and mimima of $g(r,r')$ (see the discussion below) which can be approximated by the asymptotes in Eq.~(\ref{KT}).  

\section{Numerical results: Mesoscopic system} \label{meso}

\begin{figure}[t]   
\begin{minipage}{1.2\textwidth}
   \hspace{-0.00cm}
      \mbox{\includegraphics[width=0.45\textwidth,angle=-90]{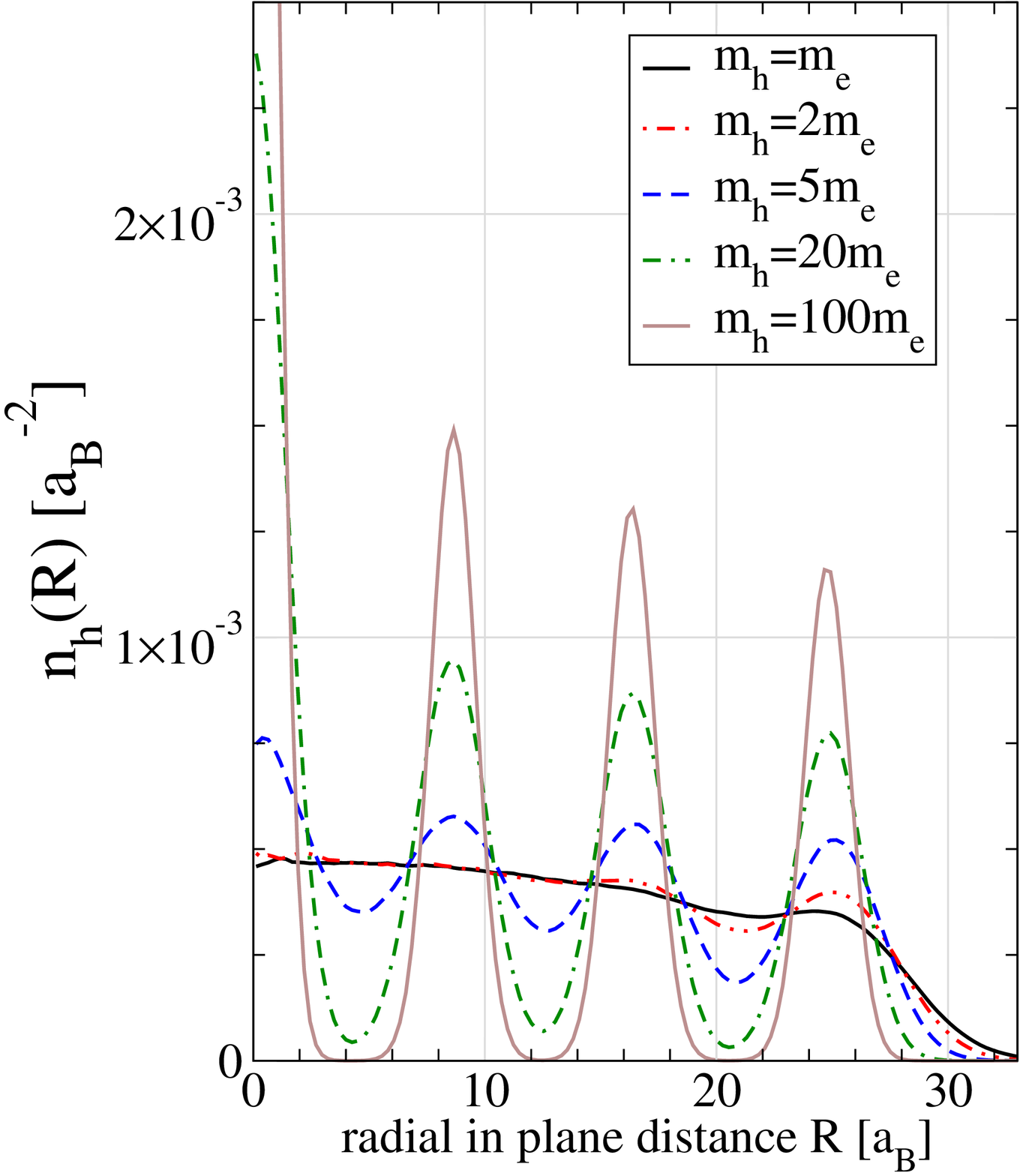}}
     \hspace{-3.3cm}
      \mbox{\includegraphics[width=0.45\textwidth,angle=-90]{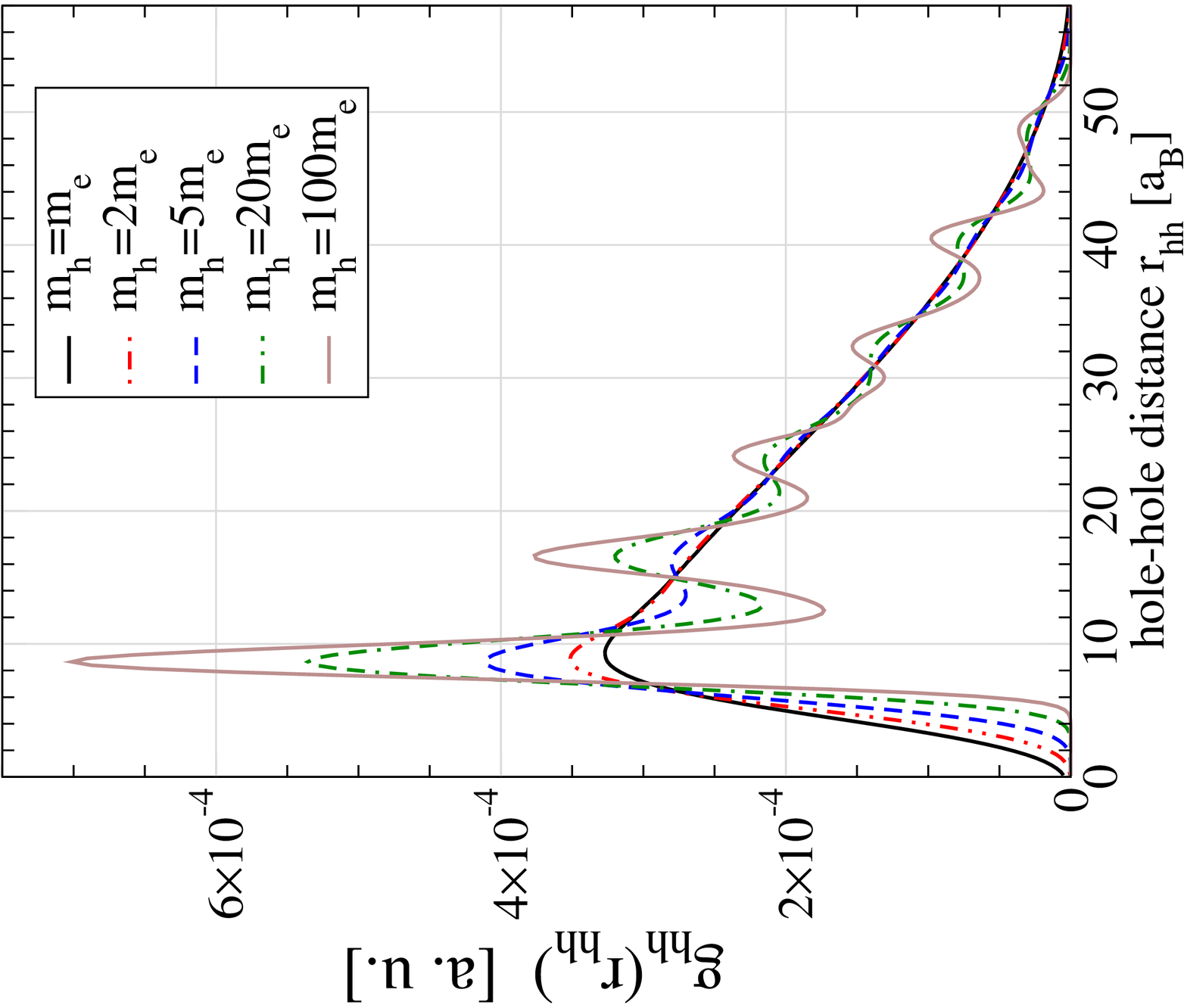}}
 \end{minipage}
\caption{(Color online) Hole radial distribution (left fig.) and hole-hole pair distribution 
(right) for $\lambda=5$ and five mass ratios (see inset) for a mesoscopic confined bilayer with $N_e=N_h=36$.
}
\label{rd_pdf}
\end{figure}

\begin{figure}[t]
\center\includegraphics[angle=-90 ,width=0.80\textwidth]{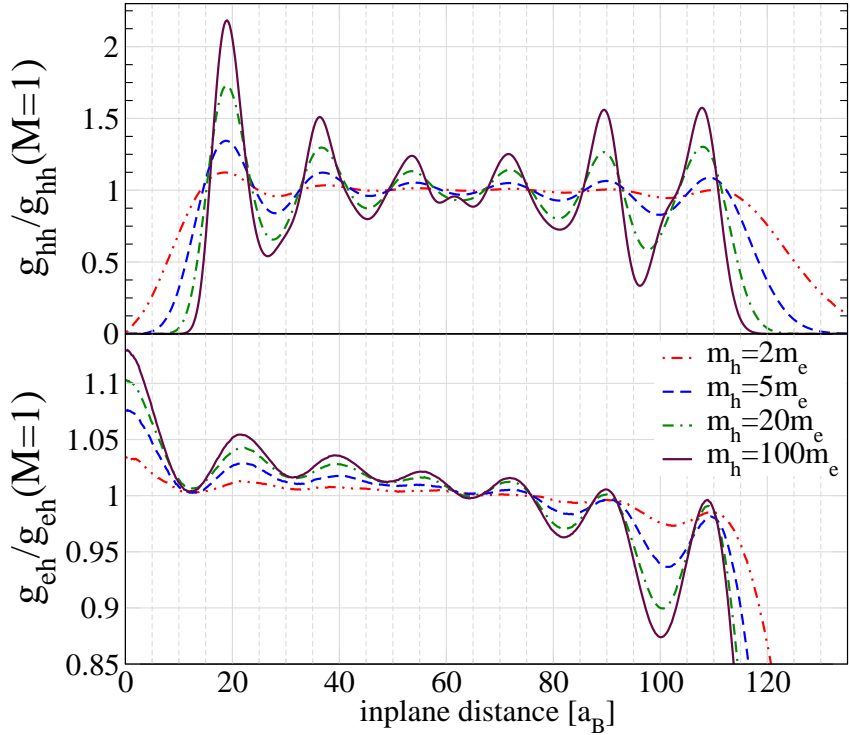}
\caption{
Hole-hole (upper fig.) and electron-hole (lower fig.) pair distribution functions for the {\em mesoscopic} bilayer with $N_e=N_h=36$ and $\lambda=10.5$ for four values of $M$ (see inset). The curves are normalized to the corresponding PDF for the case $M=1$ to eliminate the influence of the decay of the average density in the trapped system, cf. Fig.~\ref{rd_pdf}. Note that electrons and holes are always pairwise aligned vertically. The plot includes distances up to two times the radius which causes the increase of the PDF for large distances.}
\label{pairdist_meso}
\end{figure}

In the following we consider a bilayer system populated with a mesoscopic number of $N_e=N_h=36$ electrons and holes. The results of our simulations are presented in Figs.~\ref{rd_pdf}-\ref{pairdist_meso}. 
In our simulations two different densities are analyzed, given by $\lambda=5$ and $\lambda=10.5$ which corresponds to the first maxima of the pair distribution function 
$g^{max}_{hh}=8.7$ and $g^{max}_{hh}=19$, respectively. 
These densities are chosen such that, at the given temperature, hole crystallization is 
expected to occure, at least for large mass ratios $M=m_h/m_e$. If the density is chosen too low,
the Coulomb coupling would to weak for crystallization. On the other hand, if the density is 
too high, the crystal vanishes due to quantum melting.
At the chosen densities the electrons are always in the quantum liquid-like state, while the 
state of the holes can be changed by varying $M$.

At the chosen densities the total cluster radius is $R_{\lambda=5}=70a_B$ ($R_{\lambda=10.5}=150a_B$). That means that the average densities (in a single layer) are for GaAs $n_{\lambda=5}=9.4\cdot10^9/cm^2$ and $n_{\lambda=10.5}=2.0\cdot10^9/cm^2$, and for ZnSe $n_{\lambda=5}=9.9\cdot10^{10}/cm^2$ and $n_{\lambda=10.5}=2.2\cdot10^{10}/cm^2$. These values are for the electrons, 
for the holes the radius slightly decreases when $M$ is increased.

\begin{figure}[t]   
\begin{minipage}{1.2\textwidth}
  \hspace{+0.05cm}
     \mbox{\includegraphics[width=5.0cm,angle=-90]{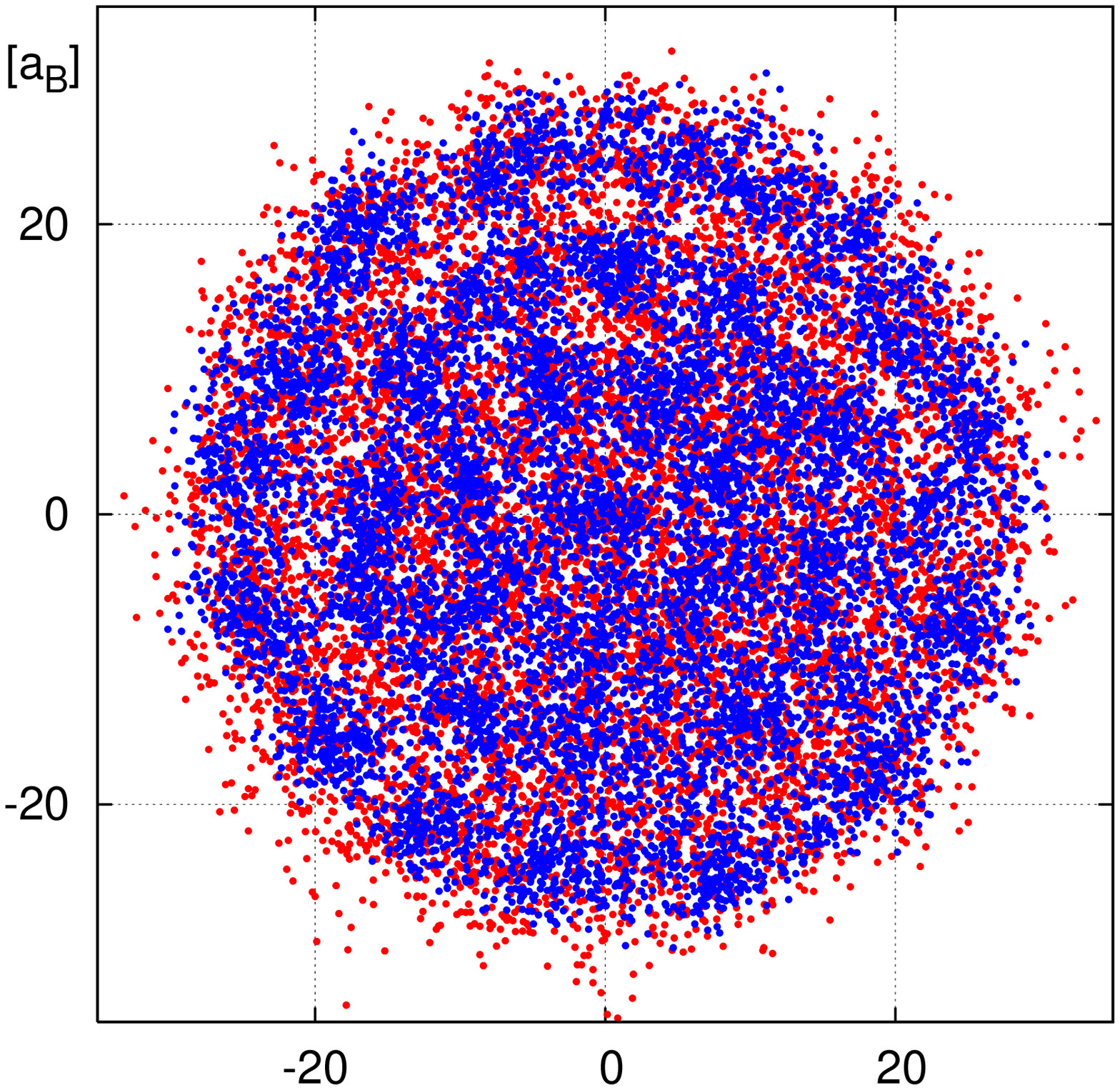}}
    \hspace{-2.08cm}
     \mbox{\includegraphics[width=5.0cm,angle=-90]{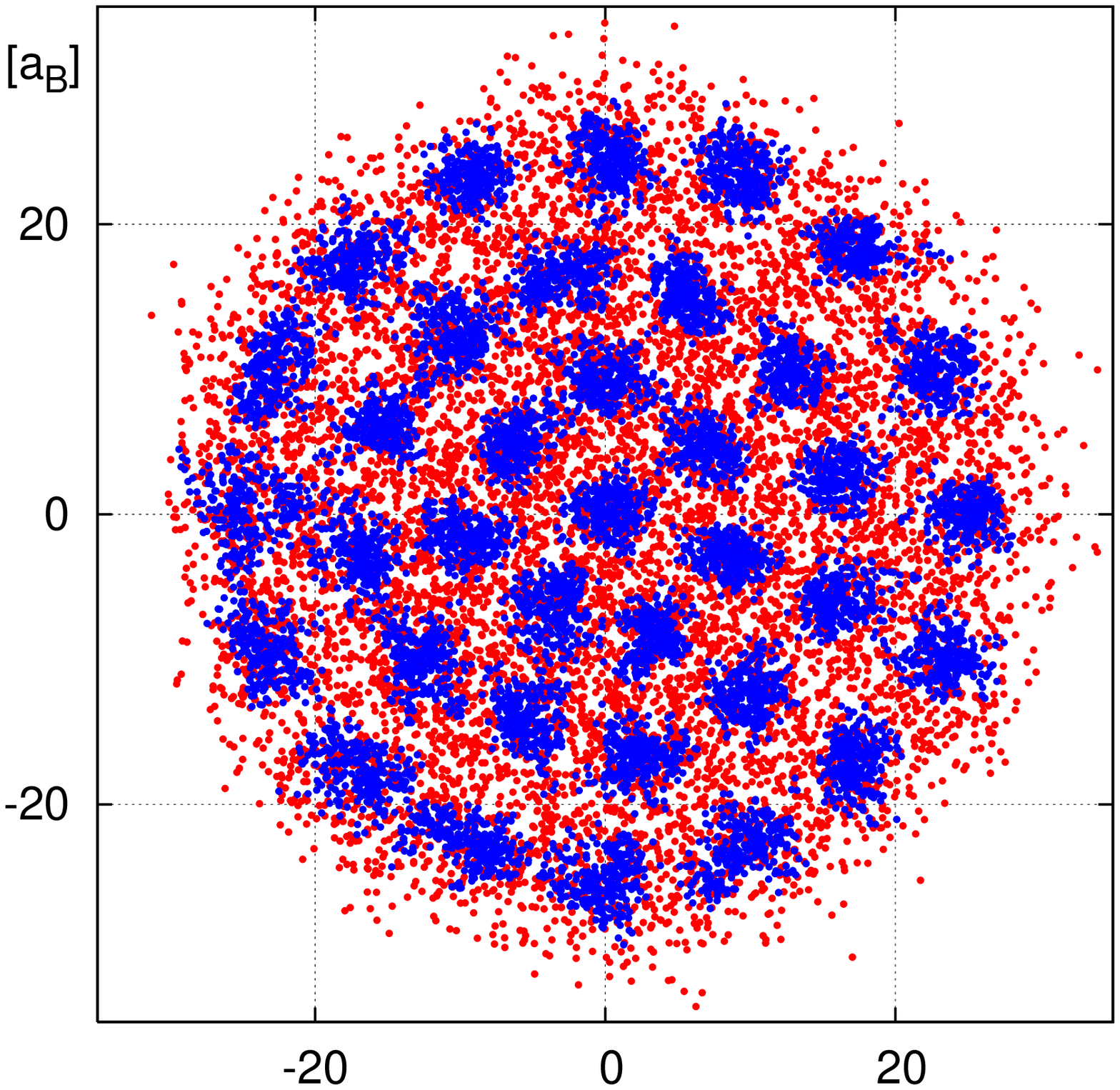}}
   \hspace{-2.07cm}
     \mbox{\includegraphics[width=5.0cm,angle=-90]{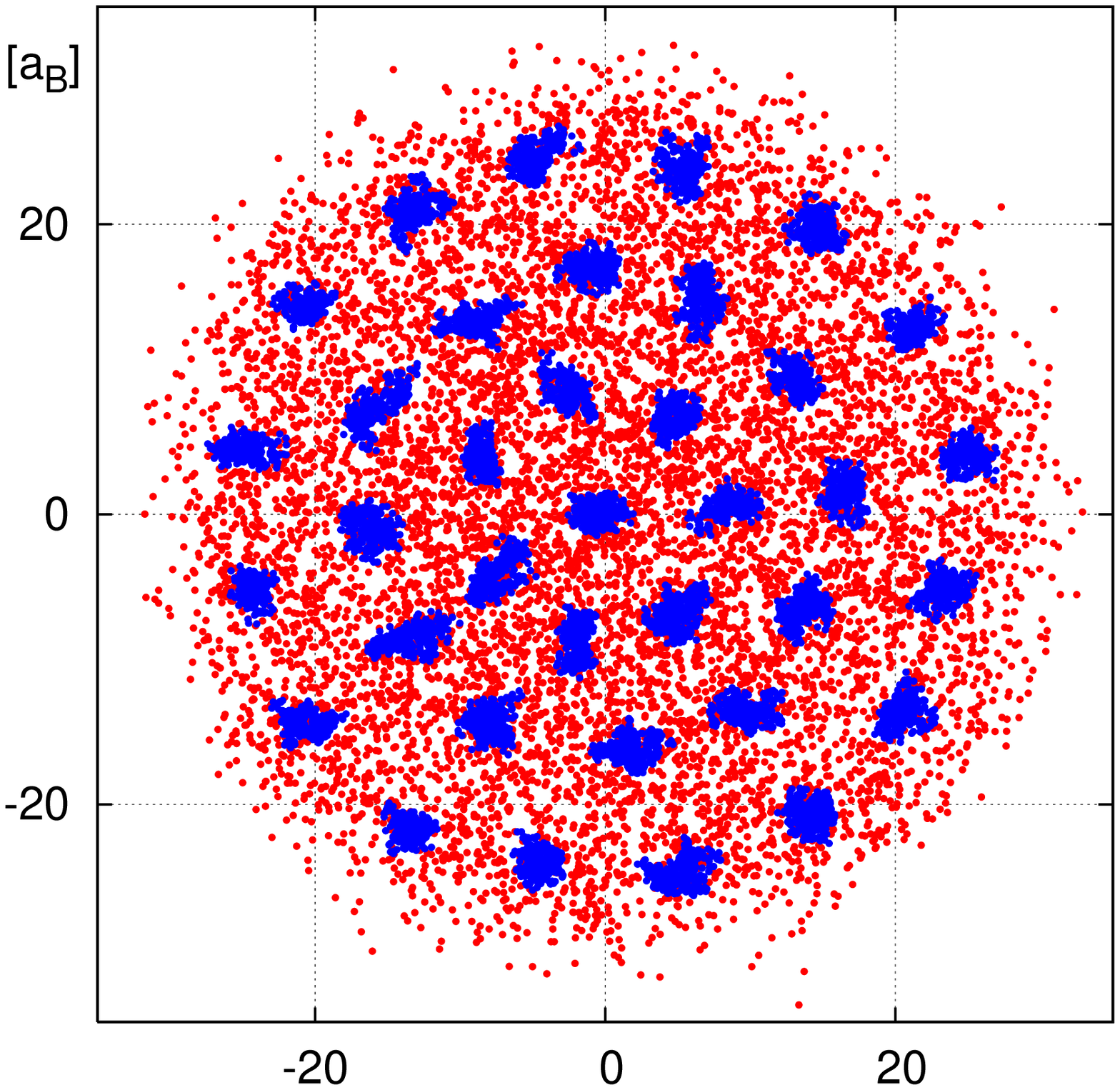}}

  \hspace{+0.05cm}
  \mbox{\includegraphics[width=5.0cm,angle=-90]{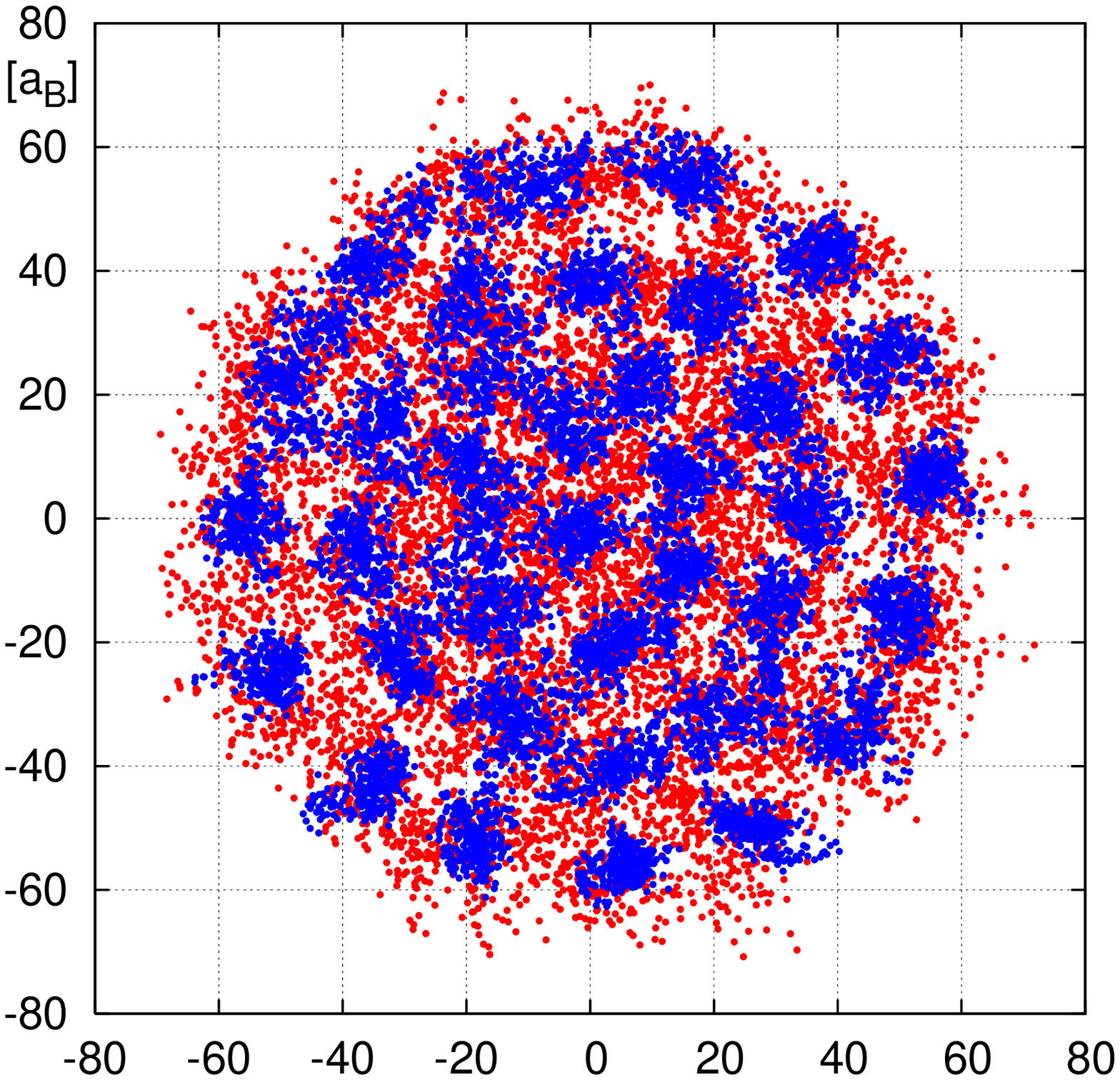}}
    \hspace{-2.08cm}
  \mbox{\includegraphics[width=5.0cm,angle=-90]{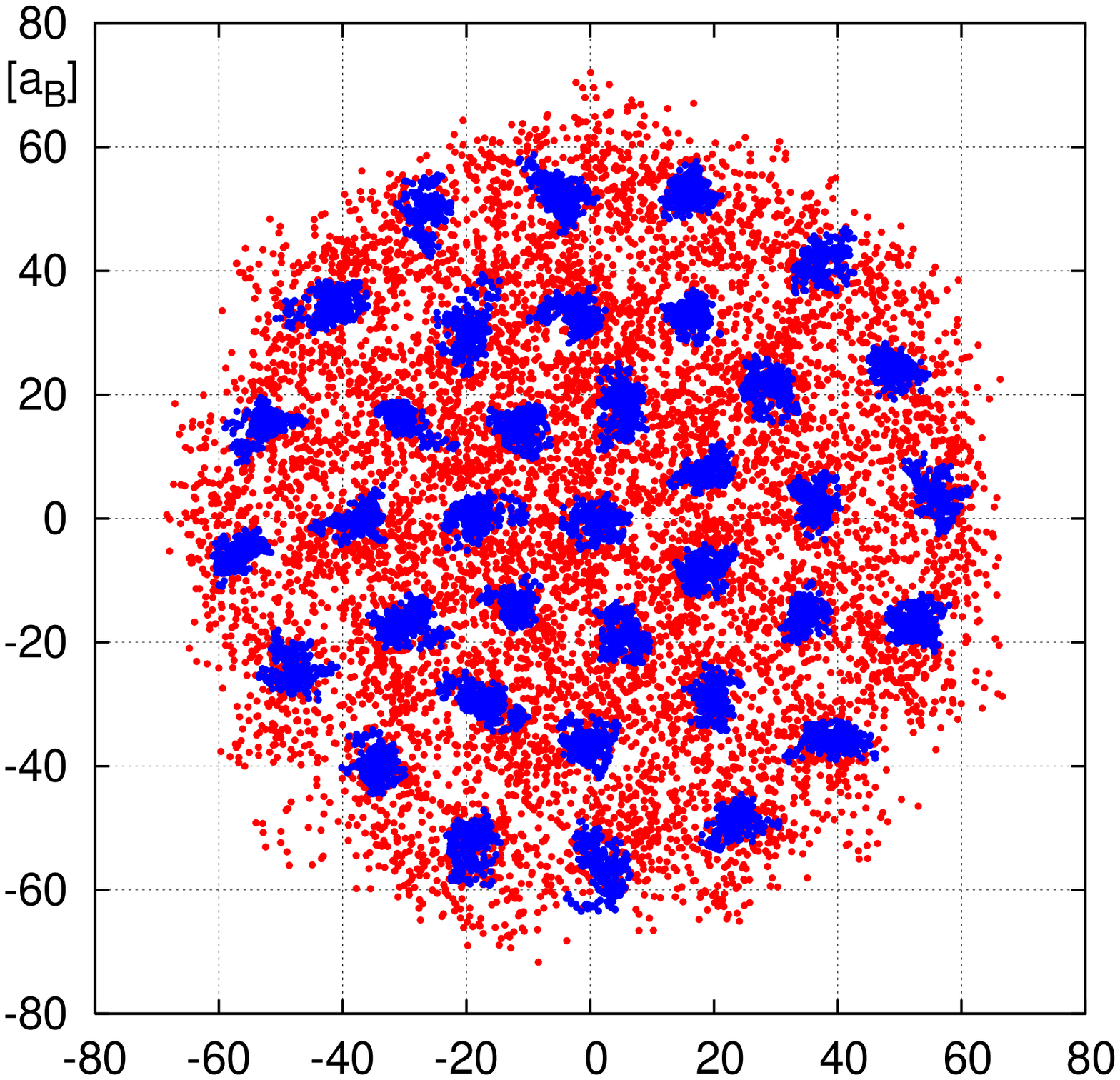}}
   \hspace{-2.07cm}
  \mbox{\includegraphics[width=5.0cm,angle=-90]{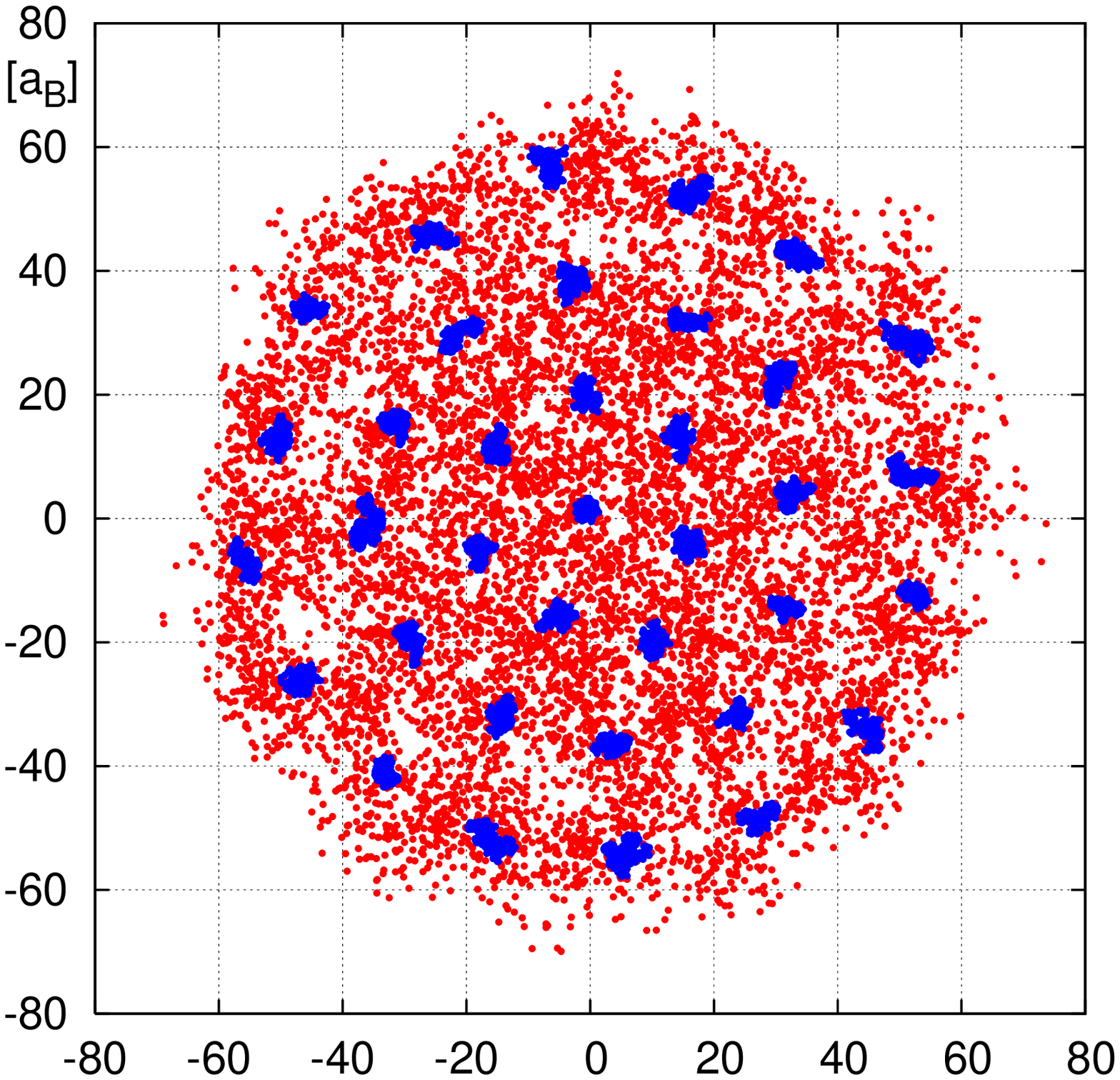}}

  \hspace{-0.00cm}
  \mbox{\includegraphics[width=4.46cm,angle=-90]{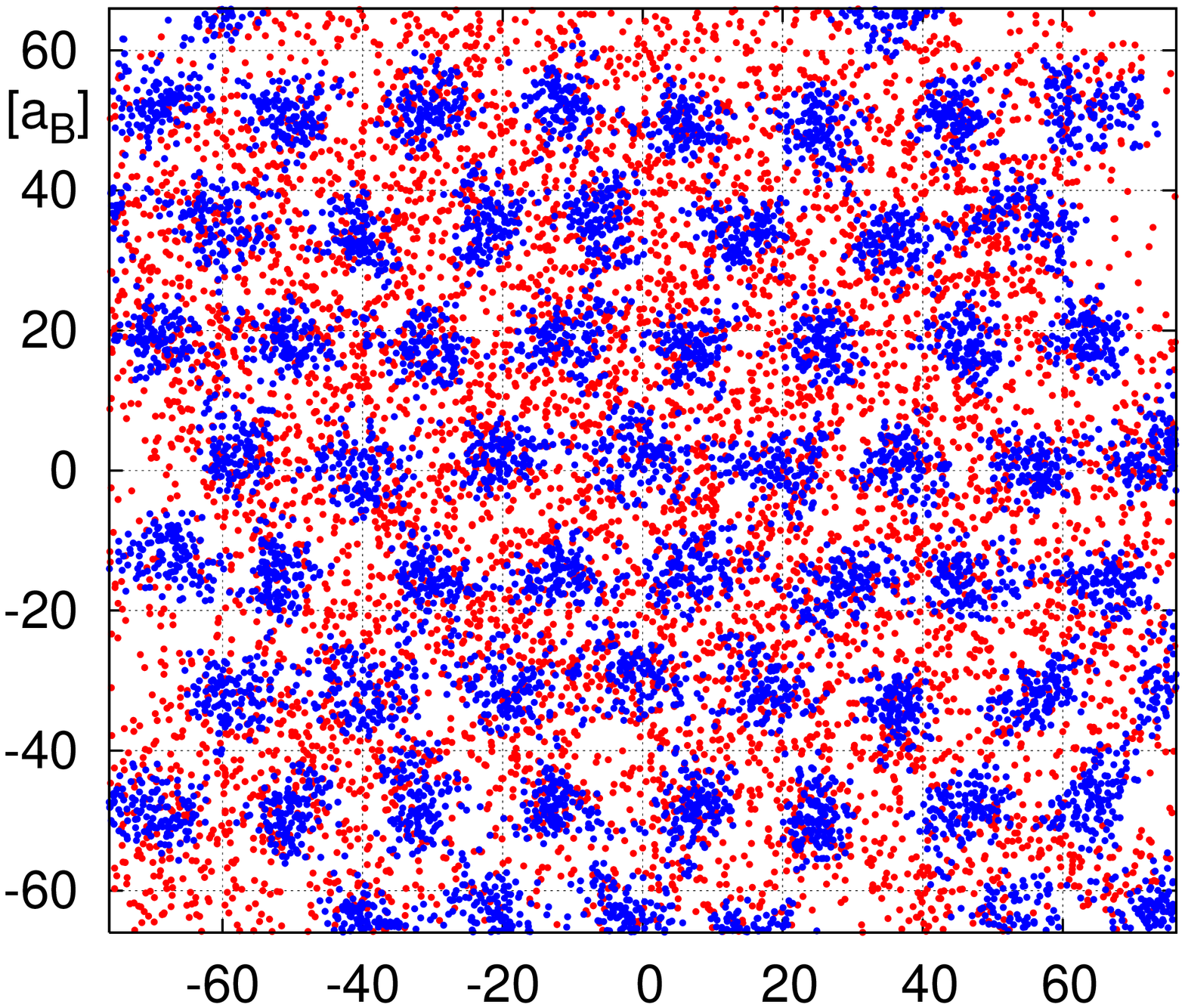}}
    \hspace{-1.3cm}
  \mbox{\includegraphics[width=4.46cm,angle=-90]{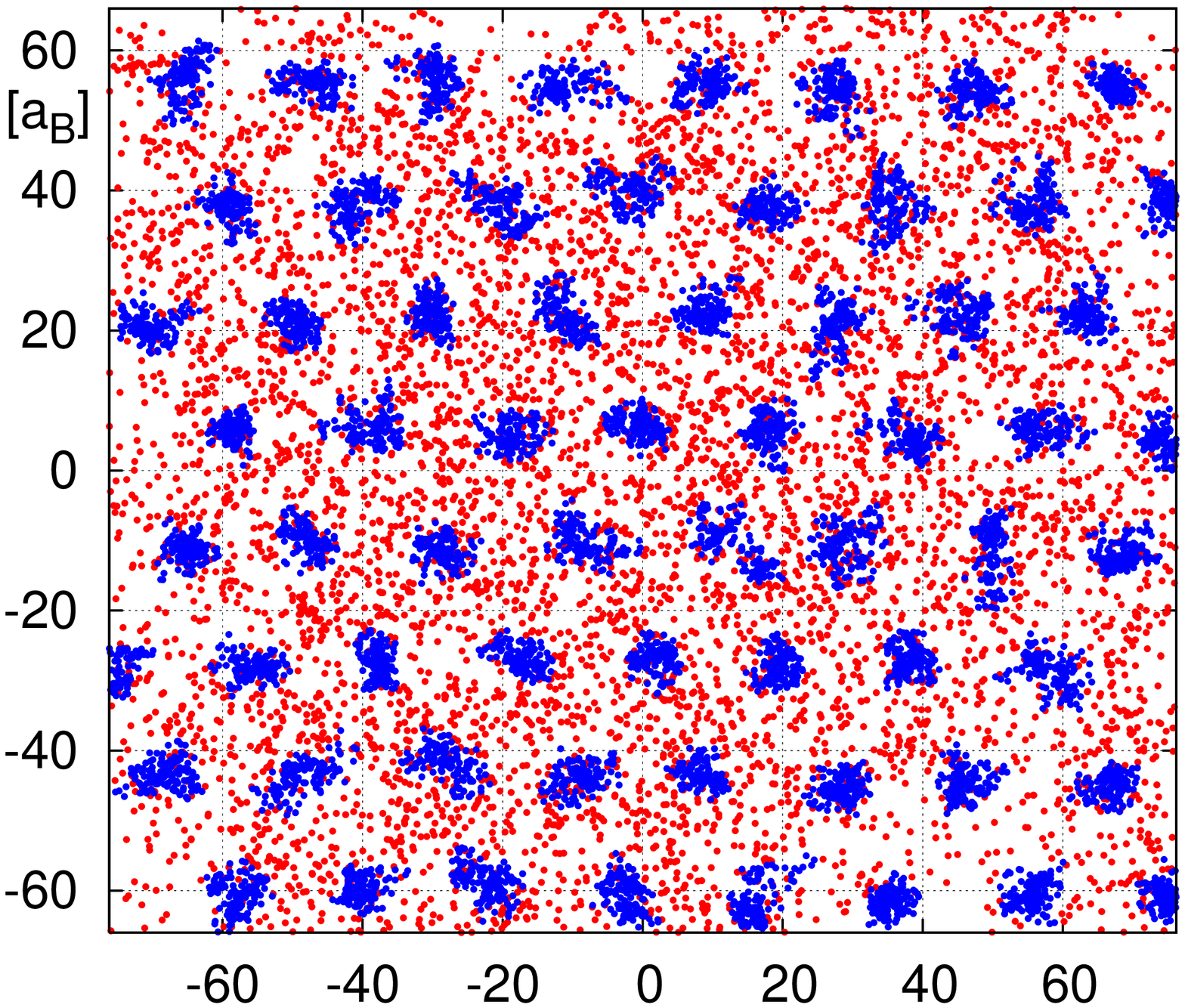}}
   \hspace{-1.3cm}
  \mbox{\includegraphics[width=4.46cm,angle=-90]{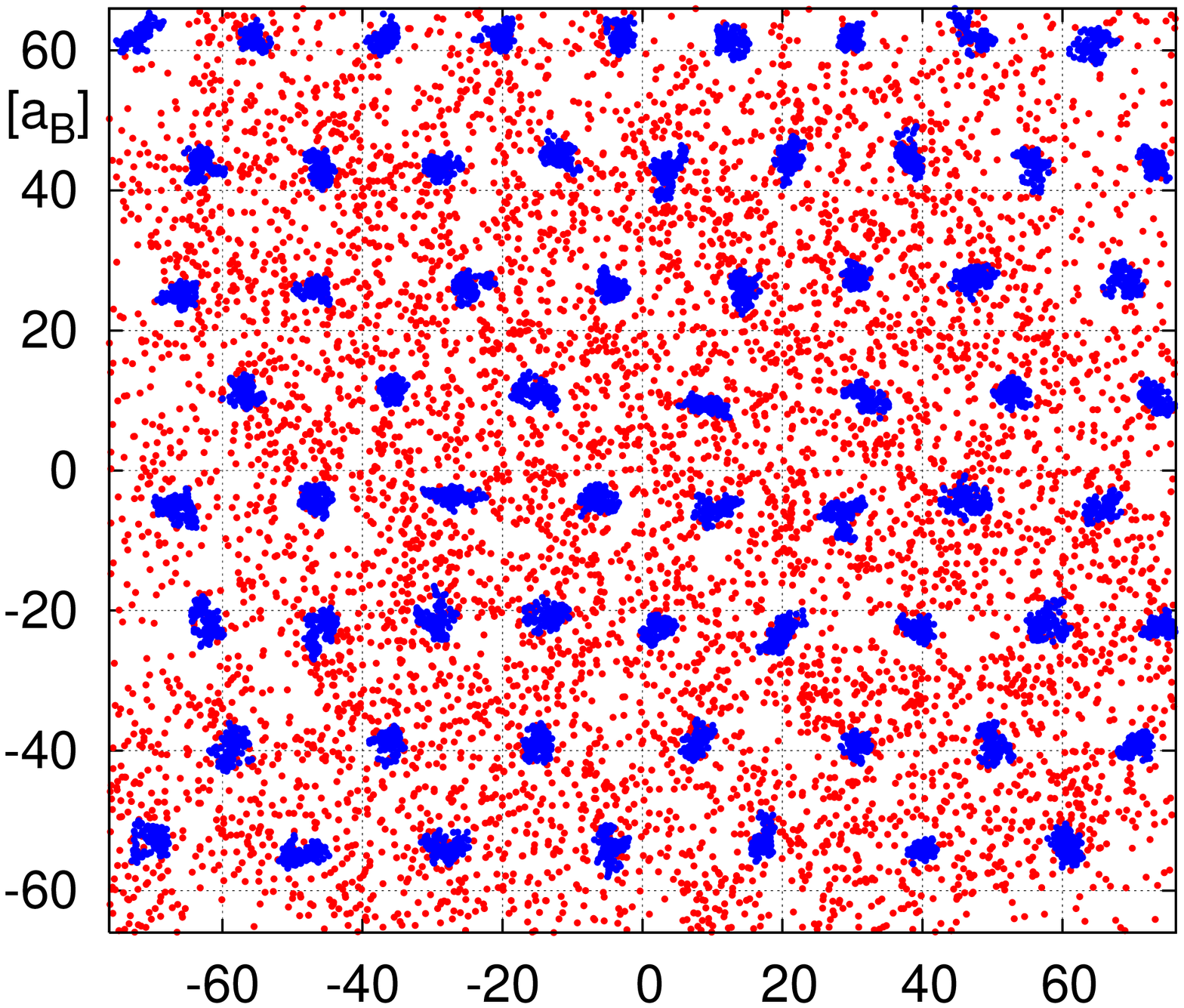}}

\end{minipage}
\caption{(Color online) Path integral Monte Carlo configuration of holes (blue points) and electrons (red dots) in a bilayer system with distance $d=20a_B$, temperature $T=Ha/3000$ and different mass ratios: $M=5$ (left column), $M=20$ (center) and $M=100$ (right). Each particle is represented 
by $256$ dots (path integral) which, for the electrons, are mutually penetrating.
{\em First two rows:} $36$ electrons and holes in a harmonic trap with coupling strengths 
$\lambda=5$ (upper panel) and $\lambda=10.5$ (second panel). Shown is a typical snapshot (without statistical averaging). Note the different axis scales in the two panels. 
{\em Row 3:} Simulation snapshots of a macroscopic bilayer ($N_e=N_h=64$ electrons and holes in the simulation cell with periodic boundary conditions, the borders mark the simulation cell. Each particle is shown only once). The density matches the one in the confined system of the second row.
There are structural defects as the triangular lettice is not uniform.}
\label{probdens}
\end{figure}

Consider first Fig.~\ref{probdens} which gives an overview on the observed behavior for the 
two densities (first two rows) when the mass ratio is varied in the range from $1$ to $100$.
The first observation is that, in all cases, the electrons are distributed almost continuously,
whereas the holes become localized when $M$ exceeds 20 (5) at $\lambda=5$ (10.5).
Due to the rotational symmetry of the trap, the holes are arranged in concentric shells.

The main difference between the mesoscopic system with a parabolic inplane confinement and an infinite system are well-known finite size effects, see e.g. ~\cite{19-20} 
which are related to the rotational symmetry instead of translational symmetry. 
Further, even when averaged over the modulation caused by the shells the density is not constant over the entire system, cf. left part of Fig.~\ref{rd_pdf}. The average density is highest in the center and decreases towards the cluster surface. 
Fig.~\ref{rd_pdf} also clearly shows the effect of the mass ratio. With increasing 
$M$ the hole-hole correlations increase leading to increased hole localiczation~\cite{AF01,Nature06}. This is accompanied by a pronounced modulation of the radial density $n(R)$ and the pair distribution (PDF) $g_{hh}$, see Fig.~\ref{rd_pdf}.
The reduction of the zero point fluctuation with increase of the particle mass $M$ leads to a hole localization and crystal formation. It is found that the shell radii in the radial density profile $n(R)$ in Fig.~\ref{rd_pdf}, as well as the peak positons in the hole-hole pair correlation function $g_{hh}$ of the mesoscopic cluster ($\lambda=5)$ are independent from mass ratio $M$. 
For $M=100$ we find that the holes are arranged in $3$ shells populated with $16$, $12$, $7$ and a single particle in the center, see Fig.~\ref{probdens}.

Fig.~\ref{rd_pdf} shows, that by changing the mass ratio from $1$ to $100$ 
the holes exhibit a transiton form a delocalized quantum state with wave function overlap to a highly ordered quasi classical state, while the electrons stay in a quantum fluid state and their correlations change only little with $M$ for the 
present parameters. We note that the classical Coulomb coupling parameter for $M=100$ is  $\Gamma_{\lambda=5}=\langle U_{corr}\rangle/\langle U_{kin}\rangle=345$ and $\Gamma_{\lambda=10.5}=158$, which is beyond the critical value for the macroscopic (OCP) crystallization $\Gamma_{crit}=137$.

Let us now consider the response of the electrons to the formation of the hole 
crystal. While the electron density is almost structurelss, some details can be 
seen in the electron-hole PDF, Fig.~\ref{pairdist_meso}. This function has a distinct 
peak at zero (in-plane) distance showing the electrons and hole are pairwise 
vertically aligned for all values of $M$. Also, the next peaks of the e-e PDF 
are aligned with those of $g_{hh}$. The small shift in the peaks of the two 
functions is due to the normalization. In order to compare the details of the 
cluster arrangements with the macroscopic system below, in Fig.~\ref{pairdist_meso} 
we have divided $g_{hh}$ and $g_{eh}$ by the corresponding functions sor $M=1$ 
where they are structureless. This allows to largely eliminate the effect of the 
trap (but slightly shifts the extrema).

\section{Numerical results: Macroscopic system}
\label{macro}

To understand the relevance of our above mesoscopic results for larger systems contaning hundreds or thousands of particles we performed additional simulations 
for a macroscopic e-h bilayer without confinement potential. 
We have considered $N_e=N_h=64$ electrons and holes in a simulation cell
of the size $\{L_x \times L_y\}=\{76.185 a_B \times 65.978 a_B\}$ with periodic boundary conditions (PBC). This corresponds to a density parameter
$r_s\approx 10$ (average particle distance in units of the electron Bohr radius). This density was chosen to be comparable to the average density in the finite system (see Sec.~\ref{meso}) for the case of coupling parameter $\lambda =10.5$.
The mass ratio $M$ was varied between $1$ and $100$, the temperature was fixed to $kT=1/3000 Ha$. 
The number of particles and the dimensions of the cell, i.e 
$L_y=\sqrt{3}L_x/2$, were choosen to best fit the symmetry of a triangular lattice. which is expected to be formed by the holes.
We note that finite size effects are of the order of few percents, a systematic 
analysis with larger particle numbers is beyond the scope of this paper.

Let us now consider the results for the macroscopic bilayer. Three typical 
shapshots for $M=5, 20, 100$ are shown in the lower row of Fig.~\ref{probdens}.
As in the mesoscopic system, for all cases the electrons are completely delocalized. 
In contrast, the hole localization increases from $M=5$ to $M=100$. Also, we 
confirm that the density of the mesoscopic system (second row) is well matched: 
the average distance between two holes as well as their extension (given by the 
size of the blue dots) is very close to the trapped case.

\begin{figure}[t]
\center\includegraphics[angle=-90 ,width=0.80\textwidth]{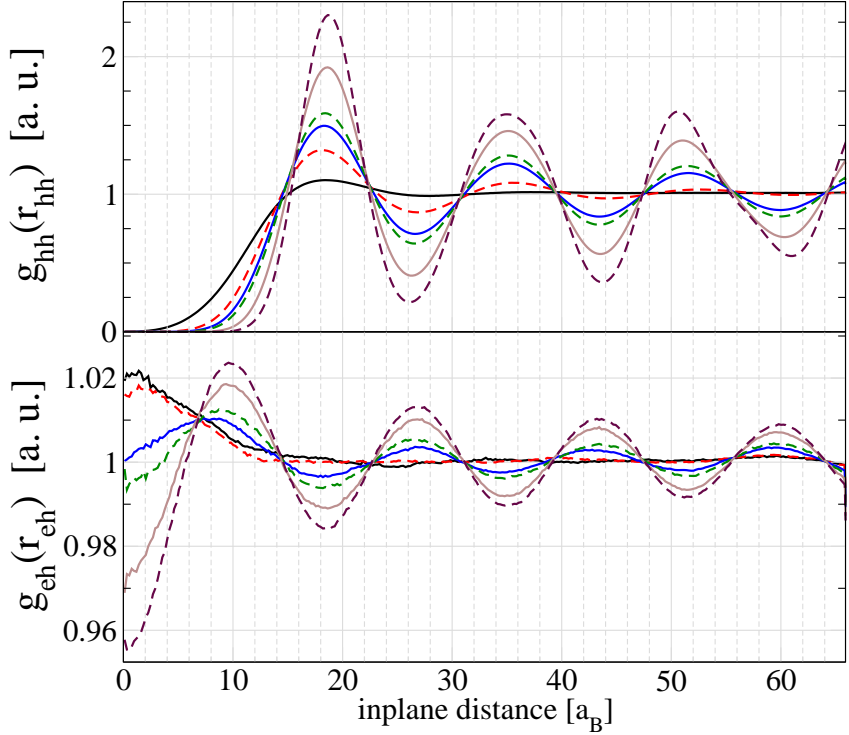}
\caption{
Hole-hole (upper fig.) and electron-hole (lower fig.) pair distribution functions for the {\em macroscopic} bilayer with $N_e=N_h=64$ (with periodic boundary conditions) for 
the mass ratios $M=1,3,4,5,10,20$ (the maxima increase with increasing $M$). 
Note the alternating location of maxima and minima of $g_{hh}$ and $g_{eh}$.
}
\label{pdf-macro}
\end{figure}

Consider now the pair distributions. In Fig.~\ref{pdf-macro} (upper fig.) we show the 
hole-hole PDF for different mass ratios $1 \leq M \leq 20$. Since the particle number 
and box size is fixed, the average particle density stays constant and the position of the first peak of the PDF are practically independent of $M$. 
However, the general behavior of the PDF changes drastically. For $M \geq 4$ we 
observe clear oscillations typical for the solid phase. Even the third and fourth peaks are well resolved (the scale exceeds half of our simulation box). These oscillations become rapidly damped by changing $M$ to $3$ and below, here
the PDF show liquid-like features. The third peak is now strongly suppressed. This 
transition can be quantified by computing the ratio of the (magnitude of the) first minimum to the first maximum which is $\gamma_1=0.48$, for $M=4$, 
and $\gamma_1=0.65$, for $M=3$. Similarly, for the third peak this ratio becomes $\gamma_3=0.76$ and $\gamma_3=0.96$, respectively. The ratio $\gamma_1$ is frequently used as an empirical criterion for the solid-liquid transition in classical systems; in a  one-component 3D system the critical value is known to be $\gamma_1^{*}\approx 1/3$. 
If a universal values exists also in the present two-component 2D quantum system where the transition is expected to be of the Kosterlitz-Thouless type is an interesting 
question which deserves further analysis. 
 
Let us now compare the pair distributions with those in the mesoscpic system at the same density ($\lambda=10.5$), Figs.~\ref{pdf-macro} and \ref{pairdist_meso}. Interestingly, we find that the first peaks of $g_{hh}$ have approximately the same height, and also the peak positons are very close, see upper parts of the two figures. 
Further we observe that the minima of $g_{hh}$ are significantly deeper in the 
macroscopic case. This is explained by intershell rotations which occur in the 
mesoscopic system \cite{AF01} and wash out the correlations. The present results 
are at temperatures above the freezout of these rotations.

Consider now the relative importance of the inter-layer correlations for the stability of the hole crystal. To this end, we have plotted the e-h PDF in Fig.~\ref{pdf-macro} (lower fig.). For the symmetric case, $M=1$, and also for $M=3$ we observe similar behavior: the highest probability has the configuration where the 
electrons reside (in their own layer) just below the holes, as was observed in the 
mesoscopic system, lower part of Fig.~\ref{pairdist_meso}. Obviously, the height
of this peak is small, the modulation depth is around $2\%$ because of the high electron degeneracy (delocalization). This means that these peaks cannot be 
associated with bound states (indirect excitons) since the electron density is well 
above the Mott density $n_{\rm Mott}$ for this system where excitons break up because the repulsion of two excitons exceeds the electron-hole binding. Note that 
$n_{\rm Mott}$ depends on the layer separation $d$ which governs the 
binding energy and the typical size $a_B^x$ of an indirect exciton which is of the order of $d$. Hence, for the present parameters, $d/a_B=20$ and $r_s\approx 10$, the in-plane exciton size exceeds the separation of two neighboring electrons which 
causes exciton ionization. On the other hand, reducing $d$ below $10$, excitons 
become stable (for temperatures below the exciton binding energy) which is 
confirmed by our PIMC simulations. 

For larger mass ratios, $M \geq 4$, a completely different behavior of $g_{eh}$
emerges. From our analysis of the hole-hole correlations we know that the holes
are now in an ``ordered state'' (or, in the terms of the Kosterlitz-Thouless theory, 
in a ``less disordered state'' with a power-low decay of off-diagonal long-range order). 
Now, there is no maximum of $g_{eh}$ at zero distance, and the function exhibits  oscillations. The explanation is that the electron density is modulated due to the 
presence of the hole crystal with maxima located in between the holes.
While the amplitude of the oscillations is small, (about $1 \%$ modulation depth) they are clearly visible and become systematically more pronounced when $M$ increases, see Fig.~\ref{pdf-macro} (lower part).
We, therefore, expect that appearance (disappearance) of these oscillations  
of $g_{eh}$ is an additional indicator of a phase transition in the present 
asymmetric bilayer system. 
  
\begin{figure}[t]   
\begin{minipage}{1.2\textwidth}
   \hspace{-0.10cm}
      \mbox{\includegraphics[width=0.38\textwidth,angle=-90]{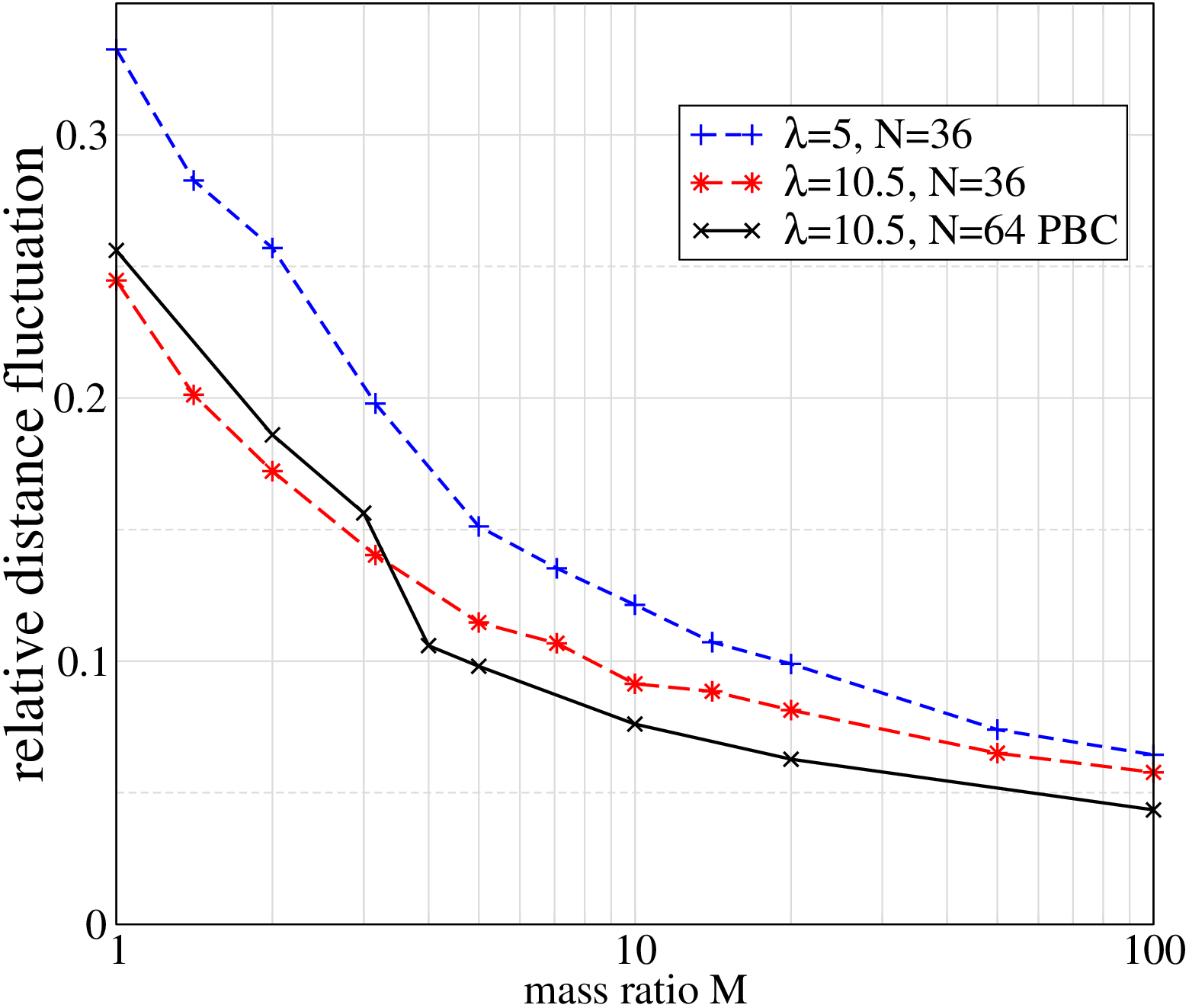}}
     \hspace{-0.9cm}
      \mbox{\includegraphics[width=0.38\textwidth,angle=-90]{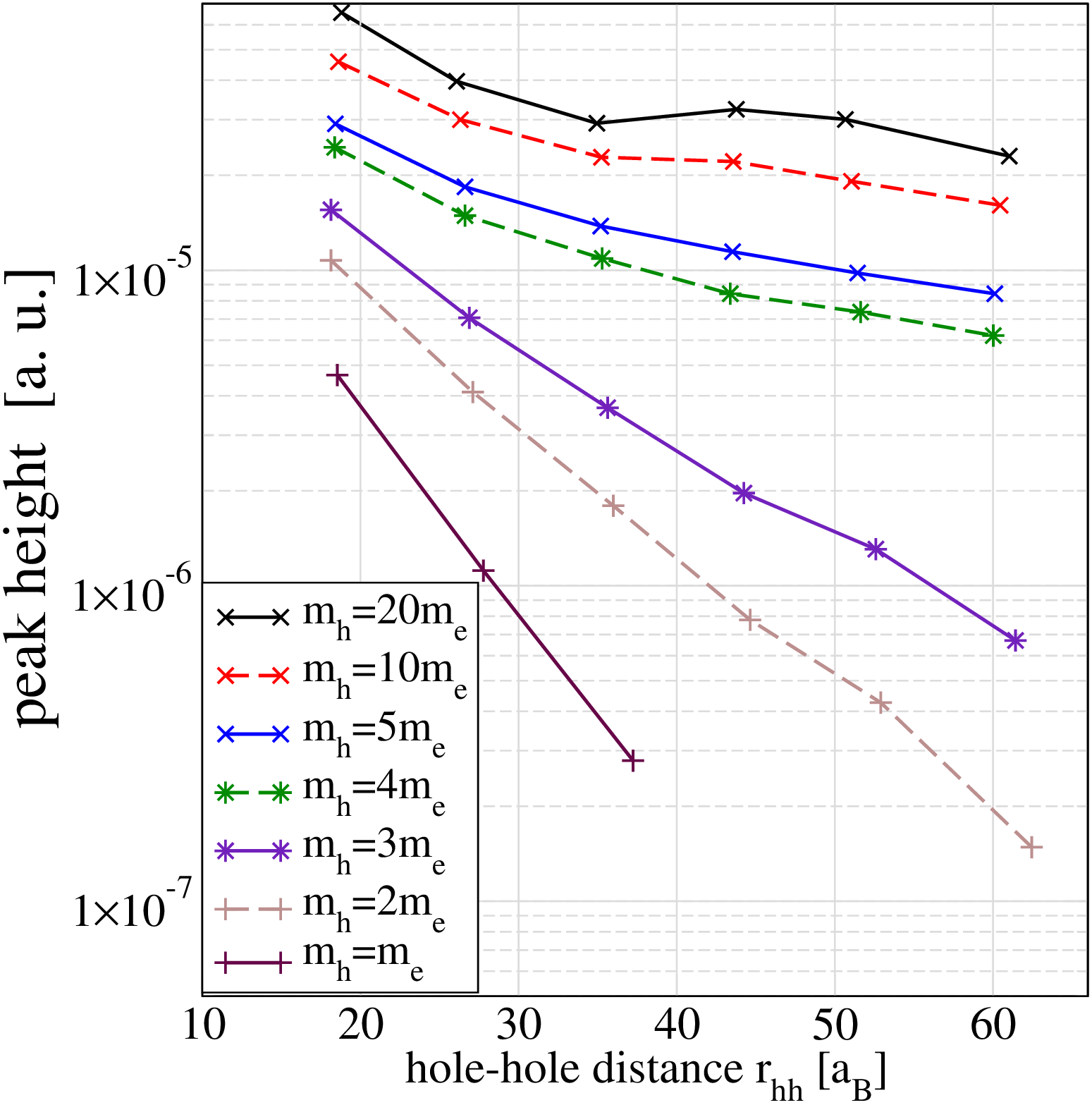}}
 \end{minipage}
\caption{(Color online) Left Fig.: Relative distance fluctuations of the holes, $u^{hh}_r$, (Eq.~\ref{reldistfluct}), as a function of the mass ratio $M$ for a macroscopic (black solid line) and mesoscopic bilayer system for two densities (see inset). 
Right Fig.: Decay of the amplitude of the maxima and minima of the hole correlations ($|g_{hh}-1|$) in the macroscopic system, cf. Fig.~\ref{pdf-macro}, for the seven mass ratios -- from 
bottom to top: $M=1,2,3,4,5,10,20$. Note the change from an exponential (for $M \leq 3$) to 
a power law decay (for $M \geq 4$) which signals the Kosterlitz-Thouless transition.}
\label{lindemann}
\end{figure}

Finally, as another quantity sensitive to phase transitions, we consider the relative distance fluctuations $u^{hh}_r$ of the holes, Eq.~\ref{reldistfluct}, as a function
of $M$, Fig.~\ref{lindemann} (right part).
This quantity exhibits a rapid drop between $M=3$ and $M=4$ which is related to 
a localization transition.
We can translate from the critical mass ratio (which is expected to be 
between three and four) to the hole density parameter $r^{(h)}_s$, using  $r^{(h)}_s=r_s^{(e)} m_h/m_e$, and the position of the first peak of the hole-hole PDF at $r_s^{(e)} \approx 10$. As a result, we obtain that the phase transition
in the hole layer occurs at a critical density in the range
$ 30 < r_s^{*(h)} < 40$. This result is close to the value $r^*_s\approx 37$ known as the critical density of solid-liquid transition in the one-component quantum 2D system at $T=0$~\cite{Cep_2D}.
Compared to this value, in our bilayer system, we observe indications of stabilization of the ``ordered state'' of the holes due to presence of the electron layer. 

We note that, at smaller values of $d$ (e.g. $d=5 a_B$ and $d=10 a_B$) no hole crystal is found. Instead we observe formation of indirect excitons which form a solid phase of composite particles. At the same time, the interparticle interaction changes
from Coulomb to dipole-like which reduces the value of the classical coupling parameter to $\Gamma=\frac{e^2 d^2}{\langle r\rangle^3}/k_BT$.
Similar tendencies have also been also in simulations of {\em symmetric} classical and quantum e-h-bilayers~\cite{Donko,Rapisarda}.    

\section{Discussion}

Analyzing the peak height (amplitude) of the $g_{hh}$ in the macroscopic system (Fig.~\ref{pdf-macro}) in dependence on the peak positon $r_{hh}$ we can deduce to the correlation decay law and compare to the asymptotics (\ref{KT}). In the disordered phase of small mass ratios $M=1...3$ we find an exponential correlation decay of  $g_{hh}$, see right part of Fig.~\ref{lindemann}. From mass ratio $M=1$ to $M=3$ disordering is lowered and the correlation length increases from $\xi = 6.5$ to $\xi = 13$. Increasing the mass ratio above the critical mass ratio, i.e. $M \geq 4$, we find a topological transition to the Kosterlitz-Thouless phase with power law correlation fall-off. 

We may now obtain a critical mass ratio at which quantum melting of the hole 
crystal takes place. Using as a criterion a critical value of $u_r=0.15$ of the 
relative hole-hole distance fluctuations we obtain $M_{crit}(\lambda=5)\approx 5$ and $M_{crit}(\lambda=10.5) \approx 2.8$, in the mesoscopic system, and 
$M_{crit}(r_s=10)\approx 3.1$, in the macroscopic system (recall that it corresponds to 
$\lambda \approx 10.5$). Obviously, the absolute numbers are 
somewhat arbitrary, but the allow for an analysis of the dominant trends. i)
$M_{crit}$ depends on density. It decreases when the coupling strength $\lambda$ 
increases in agreement with earlier observations for small e-h clusters 
\cite{HilkoCords}. ii), there is good agreement between the critical mass ratios of the mesoscopic and the macroscopic system (within $10\%$). iii) the critical values 
are much smaller than the value of $M_{crit}\approx 80$ in a 3D bulk system 
\cite{MB06} which underlines the remarkable additional control of physical 
behaviors existing in a bilayer system by a variation of the layer separation $d$.
It is expected that further reduction of $d$ will allow to further reduce $M_{crit}$ and to increase the maximum density of the hole crystal to values below $r^{(h)}_s=20$ 
\cite{AF03}.

\begin{acknowledgement}
This work has been supported by the Deutsche Forschungsgemeinschaft via SFB-TR 24, project A7 and by grants for CPU time at the Kiel Linux-Cluster ``Fermion''.
\end{acknowledgement}


\begin{thebibliography}{10}

\bibitem{kalman98} {\em Strongly Coupled Coulomb Systems}, G.~Kalman (ed.), Pergamon Press 1998

\bibitem{ions} D.J.~Wineland, J.C.~Bergquist, W.M.~Itano, J.J.~Bollinger, and
C.H.~Manney, Phys. Rev. Lett. {\bf 59}, 2935 (1987)

\bibitem{dust} H.~Thomas, G.E.~Morfill, V.~Demmel, J.~Goree, B.~Feuerbacher, and 
D.~M\"ohlmann, Phys. Rev. Lett. {\bf 73}, 652 (1994)

\bibitem{bonitz-etal.06prl} M.~Bonitz, D.~Block, O.~Arp, V.~Golubnychiy, H.~Baumgartner, P.~Ludwig, A.~Piel, and A.~Filinov, Phys. Rev. Lett. {\bf 96}, 075001 (2006)

\bibitem{AF01} A. Filinov, M. Bonitz and Yu.E. Lozovik, Phys. Rev. Lett. \textbf{86}, 3851 (2001); phys. stat. sol. (b) \textbf{221}, 231 (2000)

\bibitem{Nature06} A. Ghosal, A.~D. Guclu, C.~J. Umrigar, D. Ullmo and H. U. Baranger, Nature Phys. \textbf{2} 336 (2006)

\bibitem{MB06} M. Bonitz, V.S. Filinov, V.E. Fortov. P.R. Levashov, and H. Fehske, Phys. Rev. Lett. \textbf{95}, 235006 (2005) and J. Phys.A: Math. Gen. {\bf 39}, 4717 (2006)

\bibitem{halperin} B.I. Halperin, and T.M. Rice, Rev. Mod. Phys. {\bf 40}, 755 (1968)

\bibitem{abrikosov} A.A. Abrikosov, J. Less-Common Met. {\bf 62}, 451 (1978)

\bibitem{segretain} L. Segretain, Astron. Astrophys. {\bf 310}, 485 (1996)

\bibitem{Donko} P. Hartmann, Z. Donko and G. J. Kalman, Europhys. Lett. \textbf{72}, 396 (2005)

\bibitem{PL03} P. Ludwig, A. Filinov, M. Bonitz, and Yu.E. Lozovik, Contrib. Plasma Phys. \textbf{43}, 285 (2003)

\bibitem{PLdip} P. Ludwig, Diploma thesis: {\em Mesoscopic exciton clusters in coupled quantum dots}, Rostock University 2003 

\bibitem{Rapisarda} S.~De Palo, F.~Rapisarda and G.~Senatore, Phys. Rev. Lett. \textbf{88}, 206401 (2002); G.~Senatore and S.~De Palo, Contrib. Plasma Phys. \textbf{43}, 363 (2003)

\bibitem{AF03} A.~V. Filinov, P.~Ludwig, V.~Golubnychyi, M.~Bonitz, and Yu.E.~Lozovik, phys. stat. sol. (c) \textbf{0}, No. 5 (2003); \\
A.~V. Filinov, M.~Bonitz and Yu.~E. Lozovik, J. Phys. A: Math. Gen. \textbf{36}, 5899-5904 (2003)

\bibitem{Peeters02_PRB65} E.~Anisimovas, and F.M.~Peeters, Phys. Rev B \textbf{65}, 233302 (2002)

\bibitem{Peeters02_PRB66} E.~Anisimovas, and F.M.~Peeters, Phys. Rev B \textbf{66}, 075311 (2002)

\bibitem{Wachter} P.~Wachter, B.~Bucher, and J.~Malar, Phys. Rev. B {\bf 69}, 094502 (2004)

\bibitem{PL06} P.~Ludwig, A.~Filinov, M.~Bonitz, and H.~Stolz, phys. stat. sol. (b) \textbf{243}, 2363 (2006)

\bibitem{numbook06} A.~Filinov and M~Bonitz, in: {\em Introduction to Computational Methods for Many Body Systems}, M.~Bonitz and D.~Semkat (eds.), Rinton Press, Princeton 2006

\bibitem{storer}R.G.~Storer, J. Math. Phys. \textbf{9}, 964 (1968); A.D.~Klemm,
and R.G.~Storer, Aust. J. Phys. \textbf{26}, 43 (1973).

\bibitem{Cep95} D.M.~Ceperley, Rev. Mod. Phys. {\bf 67}, 279 (1995).

\bibitem{HilkoCords} H.~Cords, Diploma thesis: {\em Crystallization of indirect, mass-asymmetric electron-hole pairs}, Rostock University 2006

\bibitem{Cep_2D} B.~Tanatar, and D.M.~Ceperly, Phys. Rev. B {\bf 39}, 5005 (1989).

\bibitem{egger99} R.~Egger, W.~H\"ausler, C.H.~Mak, and H. Grabert,
Phys. Rev. Lett. {\bf 82}, 3320 (1999); S.~Weiss and R.~Egger, Phys. Rev. B {\bf 72}, 245301 (2005).

\bibitem{Spin_order} 
F.~Rapisarda and G.~Senatore, Aust. J. Phys. {\bf 49}, 161 (1996);
F.~Perrot and M.W.C.~Dharma-Wardana, Phys. Rev. Lett. {\bf} 87, 206404 (2001);
C.~Bulutay, B.~Tanatar, Phys. Rev. B {\bf 65}, 195116 (2002);
M.W.C.~Dharma-Wardana and F.~Perrot, Phys. Rev. Lett. {\bf 90}, 136601 (2003) and references therein. 

\bibitem{fil_prb} A.V.~Filinov, C.~Riva, F.M.~Peeters, Yu.E.~Lozovik,
and M.~Bonitz, Phys. Rev. B {\bf 70}, 035323 (2004).

\bibitem{lind} F.~Lindemann, Z.Phys {\bf 11}, 609, (1910).

\bibitem{loz_ur} V.M.~Bedanov, G.V.~Gadiyak, and Yu.E.~Lozovik,  Phys. Lett. A, 
{\bf 109} 289 (1985).

\bibitem{ber} V.L.~Berezinskii, Zh. Eksp. Theor. Fiz. {\bf 61}, 1144 (1971).

\bibitem{kost} J.M.~Kosterlitz, D.J.~Thouless, J. Phys. C
{\bf 6}, 1181 (1973).
 
\bibitem{Akopov} S.G.~Akopov and Yu.E.~Lozovik, J. Phys C: Solid State Phys. {\bf 15}, 4403 (1982).

\bibitem{Loz06} A.~Filinov, P.~Ludwig, Yu.E.~Lozovik and M.~Bonitz, to be published.

\bibitem{19-20} V.~Golubnychiy, P.~Ludwig, A.V.~Filinov, and M.~Bonitz,
Superlattices and Microstructures {\bf 34} No. 3-6, 219 (2004).


\end{thebibliography}
\end{document}